\documentstyle[aps,prd,preprint,epsfig,tighten]{revtex}
\begin{document}
\draft
\title{Effects of matter density variations on dominant oscillations\\
		in long baseline neutrino experiments}
\author{	G.L.\ Fogli, 
		G.\ Lettera,
		and E.\ Lisi\\[4mm]
}
\address{Dipartimento di Fisica and Sezione INFN di Bari\\
             	Via Amendola 173, 70126 Bari, Italy \\ }

\maketitle
\begin{abstract}
Variations around the average density and composition of the Earth mantle may
affect long-baseline (anti)neutrino oscillations through matter effects. For
baselines not exceeding a few thousand km, such effects are known to be very
small, and can be practically regarded as fractional  contributions to the
theoretical uncertainties. We perturbatively derive compact expressions to
evaluate such contributions in phenomenologically interesting scenarios with
three or four neutrinos and a dominant mass scale.
\end{abstract}
\medskip

\section{Introduction}

The increasing evidence in favor of neutrino flavor oscillations motivates
further studies of the oscillation phenomenon by means of accelerator neutrino
experiments with a (very) long baseline $L$. Current long baseline projects
include the Japanese KEK-to-Kamioka (K2K) experiment with $L=250$ km
\cite{Hi01}, the European CERN Neutrinos to Gran Sasso (CNGS) experiment  with
$L=732$ km \cite{CNGS,OPER,ICAR} (in construction),  and the American Main
Injector Neutrino Oscillation Search (MINOS) experiment  with $L=735$ km
\cite{MINO} (in construction).  Possible future projects include oscillation
searches with much longer baselines, e.g., using  $\nu$ factories
\cite{Ra01,Dy01}. Although there is still a lively debate on baseline
optimization for such future projects,  there seems to be an increasing
consensus  in favor of $L\sim 2$--$3\times 10^3$ km \cite{Dy01,NuFa}.

For baselines $L\lesssim  10^{4}$ km, neutrinos cross only the Earth mantle,
whose matter density can significantly affect the dynamics of flavor 
oscillations \cite{MSWs}. For baselines $L\lesssim 6\times 10^{3}$ km (as
considered in the present work) it has been shown extensively that the main
matter effects are accurately described by approximating the mantle density
profile with its average value along the $\nu$ trajectory  (see
\cite{Ko99,Ot01} and the extensive bibliography in \cite{Ja01}). Therefore,
variations in the matter density (and chemical composition) around the average
value can be regarded as a fractional contribution to the (theoretical) error
budget, rather than as a contribution to observable oscillation signals. It
should be also stressed that the mantle density variations around the average
value may also be affected by their own uncertainties. Indeed,  the standard
(radially symmetric) density profile provided by Preliminary Reference Earth
Model (PREM) \cite{PREM}  is subject to local variations and uncertainties at
the level of $\sim 5$--$10\%$, which may well be nonsymmetrical along a given
trajectory \cite{Ge01}.

From the above viewpoint, one would like to have a practical  recipe to
evaluate the (small) density variation effects for a variety of possible
density profiles (suggested, or at least  permitted, by local geophysical
data), without solving the $\nu$ evolution equations numerically. In this 
work, by using the perturbative approach applied in \cite{Li97} for $2\nu$
oscillations at first order in the density  variations, we  derive such a
recipe for $3\nu$ and $4\nu$ oscillation scenarios characterized by a dominant
mass scale. We think that our simple expressions may be useful to 
experimentalists involved in simulations of long baseline oscillation signals,
in order to easily quantify the effects of uncertainties associated to (known
or hypothetical) matter density variations.

\section{Notation and method}

In this section we introduce the notation and  the calculation method. The
kinematics is defined by the unitary matrix $U$ connecting flavor states
$\nu_\alpha$ with massive states $\nu_i$,
\begin{equation}
\nu_\alpha = \sum_i U_{\alpha i} \nu_i\ ,
\end{equation}
and by the squared mass matrix 
\begin{equation}
{\cal M}^2 = {\rm diag}(m^2_1,\,m^2_2,\dots)\ .
\end{equation}
The dynamics is the defined by the hamiltonian
\begin{equation}
{\cal H}(x)= U \,\frac{{\cal M}^2}{2E} \,U^\dagger +{\cal V}(x)
\end{equation}
where $\cal V$ is the matrix of potentials felt by neutrinos in matter at the
position $x$ (in the flavor basis), up to irrelevant terms proportional to the
unit matrix $\mathbf{1}$.

Solving the dynamics is equivalent to find the evolution operator $T$ from the
$\nu$ production point $x_i$ to the detection point $x_f$, 
\begin{equation}
\nu_\beta(x_f) = \sum_\alpha T_{\beta\alpha}(x_f,\,x_i)\;\nu_\alpha(x_i)\ ,
\end{equation}
which provides the desired flavor oscillation probabilities,
\begin{equation}
 P(\nu_\alpha\to\nu_\beta) \equiv P_{\alpha\beta}= |T_{\beta\alpha}|^2\ .
\end{equation}

A perturbative evaluation of $T$ can be made \cite{Li97} by splitting the $\nu$
potential matrix as
\begin{equation}
{\cal V}(x) = \overline {\cal V} +\delta {\cal V}(x)\ ,
\end{equation}
where
\begin{equation}
\overline{\cal V}=\int_{x_i}^{x_f}dx  {\cal V}(x)\Big/(x_f-x_i)\ 
\end{equation}
is the average potential matrix along the neutrino trajectory, and $\delta V$
is the residual variation, characterized by
\begin{equation}
\int_{x_i}^{x_f}dx \;\delta {\cal V}(x)=0\ .
\end{equation}
For later purposes, it is useful to distinguish variations of the neutrino
potential which are symmetric $(+)$ or antisymmetric $(-)$ with respect to the
trajectory midpoint
\begin{equation}
\overline x = (x_i+x_f)/2\ ,
\end{equation}
namely,
\begin{equation}
\delta {\cal V}(x) = \delta {\cal V}^+(x) + \delta {\cal V}^-(x)\ ,
\end{equation}
where
\begin{equation}
\delta {\cal V}^\pm  (x-\overline x) = 
\pm \delta {\cal V}^\pm(\overline x - x)\ .
\end{equation}

The hamiltonian is split as
\begin{equation}
{\cal H}(x) = \overline{\cal H} + \delta {\cal V}(x)\ ,
\end{equation}
where the constant part is
\begin{equation}
\overline {\cal H} = U\frac{{\cal M}^2}{2 E} U^\dagger+ \overline{\cal V}\ .
\end{equation}
Correspondingly, the evolution operator is  split as
\begin{equation}
T = \overline{T} + \delta T\ ,
\end{equation}
where $\overline T$ is trivially obtained by exponentiating the constant part
of $\cal H$,
\begin{equation}
\overline{T}(x_f,\,x_i) = e^{-i\overline{\cal H}(x_f-x_i)}\ ,
\label{barT}
\end{equation}
while  the correction $\delta T$ (at first order in $\delta {\cal V}$) is
perturbatively given by
\begin{equation}
\delta T (x_f,\,x_i) =  -i\int_{x_i}^{x_f}dx\; 
\overline T(x_f,\,x)\, \delta {\cal V}(x)\, \overline T(x,\,x_i) 
+O(\delta {\cal V}^2)\ .
\label{deltaT}
\end{equation}

Finally, the oscillation probability is given by 
\begin{eqnarray}
P_{\alpha\beta} &=& |\overline T_{\beta\alpha} + \delta T_{\beta\alpha}|^2\\
		&=& \overline P_{\alpha\beta} + \delta P_{\alpha\beta}\ ,
\end{eqnarray}
where
\begin{equation}
\overline P_{\alpha\beta} = |\overline T_{\beta\alpha}|^2
\label{barP}
\end{equation}
refers to the approximation of constant neutrino potential (i.e., constant
fermion density), while
\begin{equation}
\delta P_{\alpha\beta} = 2\,{\rm Re}\,
(\overline T_{\beta\alpha}\,\delta T^\ast_{\beta\alpha}) +O(\delta {\cal V}^2)
\label{deltaP}
\end{equation}
represents the (first-order) correction due to fermion density variations along
the neutrino trajectory.

In the following sections we evaluate $\delta P$ for $3\nu$ and $4\nu$
scenarios characterized (from the point of view of long baseline  experiments)
by a dominant (squared) mass scale $m^2$,  coincident with the  squared mass
difference indicated by current atmospheric neutrino experiments 
\cite{Ma01,Fo01}. Our results are valid (at first perturbative order) also
beyond  the approximation of a dominant mass scale, provided that the
subdominant oscillating terms themselves can be treated as a perturbation,
which is a good approximation in many cases of phenomenological interest for
present and future long baseline projects (see, e.g., \cite{Ce00,Fr00}).  In
this case, the effect of matter density {\em variations\/} on {\em
subdominant\/} oscillations can be regarded as a second order effect, which has
been shown to be safely negligible unless $L\gtrsim 7\times 10^3$ km
\cite{Mi01,Mi02} (a case not considered in this work).

\section{Results for three-neutrino oscillations}

Let us consider three neutrino flavors,
\begin{equation}
\nu_\alpha = (\nu_e,\,\nu_\mu,\,\nu_\tau)\ ,
\end{equation}
and a mass spectrum with two practically degenerate states,
\begin{equation}
m^2_2\simeq m^2_1\ ,
\label{deg}
\end{equation}
plus a third ``lone'' state which can be either heavier  or lighter than the
previous ``doublet'' of states,
\begin{equation}
m^2_3-m^2_{1,2} = \pm m^2\ .
\end{equation}
In the following,  the double sign $\pm$ is explicitly kept to distinguish the
so-called normal  ($+m^2$, upper sign) and inverted ($-m^2$, lower sign)
spectrum hierarchy.  The  mass spectrum can then be written as
\begin{equation}
{\cal M}^2 = 
{\rm diag}\left(\mp\frac{m^2}{2},\,\mp\frac{m^2}{2},\,\pm\frac{m^2}{2}\right)\ .
\end{equation}
The scale $m^2$ and the neutrino energy $E$ define the neutrino wavenumber
relevant for the dominant oscillations,
\begin{equation}
k=m^2/2E\ .
\end{equation}

The mixing matrix $U$ is parametrized in standard form \cite{KuPa} as
\begin{equation}
U=U_{23}(\psi)U_{13}(\phi)U_{12}(\omega)\ ,
\end{equation}
where the CP phase $\delta$ is omitted, being irrelevant under the
approximation in Eq.~(\ref{deg}).

The potential matrix can be written as
\begin{equation}
{\cal V} = {\rm diag}\left(
+\frac{V}{2},\,-\frac{V}{2},\,-\frac{V}{2}
\right)\ ,
\end{equation}
where
\begin{equation}
V(x)=\sqrt{2}\,G_F\, N_e(x)\ 
\end{equation}
and $N_e$ is the electron number density.%
\footnote{ In the Earth mantle, it is $N_e\sim 2$ mol/cm$^3$. See also Fig.~1
in  \protect\cite{Li97}.}
The above notation refers to {\em neutrinos}. The case of  {\em
antineutrinos\/} ($V\to -V$) will be discussed at the end of this section.

The constant hamiltonian $\overline {\cal H}$ is explicitly diagonalized as
\begin{equation}
\overline {\cal H}=U_{23}(\psi)U_{13}(\bar\phi)\; 
{\rm diag}({\overline E_1},{\overline E_2},{\overline E_3}) \;
U^T_{13}(\bar\phi)U^T_{23}(\psi)\ ,
\end{equation}
the energy eigenvalues being given by
\begin{eqnarray}
\overline E_1 &=& \mp \frac{\overline k}{2} \ ,\\
\overline E_2 &=& \frac{\mp k -\overline V}{2}  \ ,\\
\overline E_3 &=& \pm \frac{\overline k}{2}\ ,
\end{eqnarray}
where
\begin{equation}
s_{2\bar\phi} = \frac{s_{2\phi}}
{\sqrt{(c_{2\phi}\mp \overline V/k)^2+s^2_{2\phi}}}
\end{equation}
gives the neutrino mixing angle $\bar\phi$ in matter ($c=\cos$, $s=\sin$), and
\begin{equation}
\overline k = k \frac{s_{2\phi}}{s_{2\bar\phi}}
\end{equation}
is the neutrino wavenumber in matter (for average electron density).

By using Eqs.~(\ref{barT}) and (\ref{barP}) one gets 
\begin{eqnarray}
\left(
\begin{array}{ccc}
\overline P_{ee} & \overline P_{e\mu} & \overline P_{e\tau}\\
\overline P_{\mu e} & \overline P_{\mu\mu} & \overline P_{\mu\tau}\\
\overline P_{\tau e} & \overline P_{\tau\mu} & \overline P_{\tau\tau}
\end{array}
\right) = {\mathbf 1} &+& 
\left(
\begin{array}{ccc}
0 & 0 & 0 \\
0 & -1 & 1 \\
0 & 1 & -1
\end{array}
\right) s^2_{2\psi} (c^2_{\bar\phi}\;S^2_{23}+s^2_{\bar\phi}\;S^2_{12})
\nonumber \\
&+&
\left(
\begin{array}{ccc}
-1 & s^2_\psi & c^2_\psi\\
s^2_\psi & -s^4_\psi & -s^2_\psi c^2_\psi \\
c^2_\psi & -s^2_\psi c^2_\psi & -c^4_\psi
\end{array}
\right) s^2_{2\bar\phi}\;S^2_{31}
\label{barP3}
\end{eqnarray}
for the oscillation probabilities with constant (average) potential $\overline
V$, where the oscillation factors are given by
\begin{equation}
S^2_{jl}=\sin^2\left(\frac{\overline E_j-
\overline E_l}{2}(x_f-x_i)\right)\ .
\end{equation}
The above results for $\overline P_{\alpha\beta}$  are well-known (see, e.g.,
\cite{Fo97}).

We then compute the desired  corrections $\delta P_{\alpha\beta}$ due to
electron density variations through Eq.~(\ref{deltaP}). Omitting the algebra,
the final results read
\begin{eqnarray}
\left(
\begin{array}{ccc}
\delta P_{ee} & \delta P_{e\mu} & \delta P_{e\tau}\\
\delta P_{\mu e} & \delta P_{\mu\mu} & \delta P_{\mu\tau}\\
\delta P_{\tau e} & \delta P_{\tau\mu} & \delta P_{\tau\tau}
\end{array}
\right) =   & &  
\left(
\begin{array}{ccc}
0 & 0 & 0 \\
0 & -1 & 1 \\
0 & 1 & -1
\end{array}
\right) s^2_{2\bar\phi}s^2_\psi c^2_\psi \;C\,S'
\nonumber \\
&+&
\left(
\begin{array}{ccc}
-1 & s^2_\psi & c^2_\psi\\
s^2_\psi & -s^4_\psi & -s^2_\psi c^2_\psi \\
c^2_\psi & -s^2_\psi c^2_\psi & -c^4_\psi
\end{array}
\right) s^2_{2\bar\phi}c_{2\bar\phi}\;C\, S''\;\ ,
\label{deltaP3}
\end{eqnarray}
where $S'$ and $S''$ are oscillating terms defined as
\begin{eqnarray}
S' &=& \sin\left(\frac{\mp k -\overline V}{2}(x_f-x_i)\right)\ ,\\
S'' &=& \sin\left(\pm\frac{\overline k}{2}(x_f-x_i)\right)\ ,
\end{eqnarray}
while $C$ is basically the Fourier (cosine) transform of the symmetric part of
the potential variations,
\begin{equation}
C(\overline k)=
\int_{x_i}^{x_f}dx\;\delta V^{+}(x)\cos\left(\overline k(x-\overline x)\right)
\ .
\label{Fou}
\end{equation}

By comparing Eqs.~(\ref{barP3}) and (\ref{deltaP3}), it emerges that the
effects of nonconstant electron density  can be simply embedded as corrections
to the oscillating factors $S^2_{jl}$,
\begin{equation}
S^2_{jl} \longrightarrow S^2_{jl} + C \times
\left\{
\begin{array}{ll}s^2_{\bar\phi}c^2_{\bar\phi}\;S'
& {\rm if\ }jl=23{\rm\ or\ }12\ ,\\
c_{2\bar\phi}\;S''
& {\rm if\ }jl=13\ .
\end{array}
\right.
\label{Rep3}
\end{equation}

The above replacement represent our simple recipe for the evaluation of matter
density variations in the $3\nu$ scenario. Once the usual ``constant density''
probabilities [Eq.~(\ref{barP3})] have been computed for a given long baseline
experiment configuration, one has just to calculate the Fourier term $C$
[Eq.~(\ref{Fou})] and make the substitutions in Eq.~(\ref{Rep3}) to get the
corrected probabilities. By using a variety of possible density profiles
(allowed or suggested by geophysical information along the given baseline) one
can then  easily evaluate the effect of uncertainties in $N_e$ upon observable
quantities, without solving numerically the  neutrino evolution equations for
the specific profile.

Several remarks are in order, about the size and the properties of the
correction terms $\delta P_{\alpha\beta}$  in Eq.~(\ref{deltaP3}). Their size
is typically rather small $(\lesssim 10^{-3})$,  both because they are
suppressed by an overall factor $\sin^2 2\bar\phi$  (constrained to be
$\lesssim 0.1$ by reactor%
\footnote{Although reactor bounds apply, strictly speaking, to $\phi$ and not
to $\bar\phi$, our arguments are not qualitatively changed.}
and atmospheric neutrino data \cite{Ma01}), and because the integral $C$ is
typically rather small, due to the oscillating behavior of the integrand. The
maximum of $C$ is reached only for a hypothetical configuration of the matter
profile  with $\delta N_e^+(x) \propto \cos[\overline k(x-\overline x)]$,
namely, when matter density fluctuations happen to be ``in phase'' with 
oscillations. In such a contrived case, since
\begin{equation}
dx\,\delta V^+ = 0.386 \times 10^{-3}\left(\frac{dx}{\rm km}\right) 
\left(\frac{\delta N^+_e}{{\rm mol}/{\rm cm}^3} \right)
\end{equation}
with (plausibly)   $\delta N_e \lesssim 10\% N_e \simeq 0.2$ mol/cm$^3$, the
integral $C$ could be as large as a few percent over long baselines ($\gtrsim
10^3$ km). However, the realization of such a ``maximum effect'' configuration
(if approximately reachable) can be attained only in a  narrow range of $k$
(i.e., of neutrino energy). Outside such a range, the size of $C$  (and thus of
$\delta P_{\alpha\beta}$) is rapidly suppressed. We refer the reader to the
recent papers \cite{Sh01,Ja01} for extensive  numerical investigations of the
size of $\delta P_{\alpha\beta}$,  for various admissible $\delta N_e(x)$
profiles.

Concerning the symmetry properties of the oscillation probabilities, notice
that the first-order corrections $\delta P_{\alpha\beta}$ do not involve the
antisymmetric part of the potential. As a consequence, $T$-violation effects
are absent not only at 0th order, but also at 1st order in $\delta V$,
\begin{eqnarray}
\overline P_{\alpha\beta} &=& \overline P_{\beta\alpha}\ ,\\
P_{\alpha\beta} &=& P_{\beta\alpha} + O(\delta V^2)\ ,
\end{eqnarray}
in agreement with the discussion in \cite{Mi01,Mi02}.

Finally, let us consider the case of {\em antineutrinos}, for which the 
potential is given by $-V$. It is easy to check that all the above oscillation
probabilities are invariant under the simultaneous replacements $\pm m^2\to \mp
m^2$ (change of hierarchy) {\em and\/} $+V\to -V$ (change of potential).
Therefore, antineutrino probabilities for direct (inverse) hierarchy are equal
to neutrino probabilities for inverse (direct) hierarchy,
\begin{equation}
P(\pm m^2;\,\overline\nu_\alpha\to\overline\nu_\beta)=
P(\mp m^2;\,\nu_\beta\to\nu_\alpha)\ .
\end{equation}

\section{results for four-neutrino oscillations}

In this section we consider the phenomenologically interesting scenario of
three active plus one sterile neutrino $\nu_s$, with a mass spectrum consisting
of two separated doublets (see \cite{Fo01} and references therein). In such a
case, it is useful to decompose the flavor space as
\begin{equation}
(\nu_e,\,\nu_\mu,\,\nu_\tau,\,\nu_s)=(\nu_+,\,\nu_\mu)\oplus(\nu_-,\,\nu_e)
\end{equation}
where the states $\nu_\pm$ are linear combinations of $\nu_\tau$ and $\nu_s$,
\begin{equation}
\left(
\begin{array}{c}
\nu_+ \\
\nu_- 
\end{array}
\right)=
\left(
\begin{array}{cc}
c_\xi & s_\xi\\
-s_\xi & c_\xi 
\end{array}
\right)
\left(
\begin{array}{c}
\nu_\tau \\ \nu_s
\end{array}
\right)\ .
\end{equation}
Under plausible approximations discussed in \cite{Fo01}, the mass-mixing
parameter space can then be factorized as
\begin{equation}
4\nu {\rm\ parameter\ space} \simeq  (m^2,\,\psi,\,\xi)\otimes
(\delta m^2,\,\omega,\,\xi)\ ,
\end{equation}
where the first and second factors dominate the oscillation physics in
terrestrial and solar experiments, respectively. From the point of view of
long-baseline experiments, the role of $(m^2,\psi)$ is similar to the $3\nu$
case in the previous section, while the new angle $\xi$ interpolates smoothly
from pure active oscillations ($s_\xi=0$) to pure sterile oscillations
($s_\xi=1$).

In the $(\nu_+,\nu_\mu)$ basis, the effective hamiltonian for long baseline
experiments is given by  \cite{Fo01}
\begin{equation}
\overline {\cal H} = \frac{1}{2} 
\left(
\begin{array}{cc}
\overline V-k\,c_{2\psi} & k\,s_{2\psi}\\
k\,s_{2\psi} & -\overline V+k\,c_{2\psi}
\end{array}
\right)\ ,
\end{equation}
where the average  neutrino potential is now related to the average neutron
density $\overline N_n$,
\begin{equation}
\overline V = \sqrt{2}\, G_F\, s^2_\xi\, {\overline N}_n/2\ .  
\end{equation}

The dynamics is then effectively reduced to an equivalent two-flavor problem,
which is easily solved. We give the final results for the relevant muon
neutrino probabilities,
\begin{eqnarray}
\overline P_{\mu\mu}  &=& 1-s^2_{2\bar\psi}\,S^2\ ,\\
\overline P_{\mu\tau} &=& c^2_\xi\,s^2_{2\bar\psi}\,S^2\ ,\\
\overline P_{\mu s}   &=& s^2_\xi\,s^2_{2\bar\psi}\,S^2\ ,
\end{eqnarray}
where
\begin{equation}
s_{2\bar\psi} = \frac{s_{2\psi}}{\sqrt{(c_{2\psi}-\overline V/k)^2+s^2_{2\psi}}}
\end{equation}
and the dominant oscillating factor is
\begin{equation}
S^2=\sin^2\left (\overline \frac{k}{2} (x_f-x_i)\right)\ ,
\end{equation}
with
\begin{equation}
\overline k = k \frac{s_{2\psi}}{s_{2\bar\psi}}\ .
\end{equation}

We perturbatively find that the first-order correction due to nonconstant
neutron density amounts to the following replacement
\begin{equation}
S^2 \to S^2 + c_{2\bar\psi}\, C\, \sin\left( 
\frac{\overline k}{2}(x_f-x_i) \right )\ ,
\label{Rep4}
\end{equation}
where $C$ is formally defined as in Eq.~(\ref {Fou}).

Notice that, contrary to the previous $3\nu$ scenario, the cases $+m^2$ and
$-m^2$ are now equivalent, being connected by the replacement $\psi \to\
\pi/2-\psi$. One can then fix the sign of $m^2$, provided that $\psi$ is taken
in its full range $[0,\pi/2]$.

Finally, antineutrino probabilities are obtained by the above neutrino
probabilities through the replacement $V\to -V$ or, equivalently, through
$m^2\to -m^2$ or, equivalently, through $s_\psi \to c_\psi$.

\section{summary}

In the context of long baseline $\nu$ oscillation experiments, the small
effects due to nonconstant matter density (along a given neutrino trajectory)
can be regarded as a fractional contribution to the uncertanties associated to
oscillation signals. Therefore, it makes sense to search for a practical method
to compute them for given density profiles. We perturbatively find that the
first-order effects can be simply embedded as a correction to the oscillating
factors in both $3\nu$ and $4\nu$ scenarios with a dominant mass scale
[Eqs.~(\ref{Rep3}) and (\ref{Rep4}), respectively]. The correction involves the
Fourier (cosine) transform of the symmetric part of the density variations,
which can be easily evaluated and incorporated in experimental simulations.

\acknowledgments

This work was partly supported by the Italian  {\em Istituto Nazionale di
Fisica Nucleare\/} (INFN) and  {\em Ministero dell'Istruzione,
dell'Universit\`a e della Ricerca\/}  (MIUR) under the ``Astroparticle
Physics'' project.


\eject

\begin{thebibliography}{99}


\bibitem{Hi01}	J.E.\ Hill for the K2K Collaboration, in the Proceedings of
		{\em Snowmass~'01}, Summer Study on the Future of Particle
		Physics (Snowmass, Colorado, 2001), hep-ex/0110034. The K2K
		experiment is currently interrupted, pending reinstrumentation
		of the Super-Kamiokande detector.

\bibitem{CNGS}	See the CNGS project website, 
		proj-cngs.web.cern.ch/proj-cngs~.

\bibitem{OPER}	OPERA Collaboration, M.\ Guler {\em et al.}, CERN Report
		SPSC-2001-025, available at operaweb.cern.ch~.

\bibitem{ICAR}	A.\ Rubbia in {\em Neutrino 2000}, 19th International
		Conference on Neutrino Physics and Astrophysics
		(Sudbury, Canada, 2000), Nucl.\ Phys.\ B (Proc.\ Suppl.)
		{\bf 91}, 223 (2000); see also the ICARUS website
		www.aquila.infn.it/icarus~.

\bibitem{MINO}	S.G.\ Wojcicki 
		in {\em Neutrino 2000} \protect\cite{ICAR}, p.~222;
		see also the MINOS website www-numi.fnal.gov:8875~.

\bibitem{Ra01}	R.\ Raja {\em et al.}, hep-ex/0108041.

\bibitem{Dy01}	F.\ Dydak, in the Proceedings of {\em TAUP~'01\/},
		7th International Workshop on Topics in Astroparticle and
		Underground Physics (Laboratori Nazionali del Gran Sasso,
		Italy, 2001), to appear.

\bibitem{NuFa}	Proceedings of {\em NuFact~'01}, 3rd International
		Workshop on Neutrino Factory Based on Muon Storage Rings
		(Tsukuba, Japan, 2001), to appear; transparencies available
		at psux1.kek.jp/$^\sim$nufact01~. See, in particular, the
		theoretical summary talk by O.\ Yasuda, hep-ph/0111172.

\bibitem{MSWs}	L.\ Wolfenstein, Phys.\ Rev.\ D {\bf 17}, 2369 (1978);
		S.P.\ Mikheyev and A.Yu.\ Smirnov, Yad.\ Fiz.\ {\bf 42},
		1441 (1985) [Sov.\ J.\ Nucl.\ Phys.\ {\bf 42}, 913 (1985)];
		Nuovo Cimento C {\bf 9}, 17 (1986);
		V.\ Barger, S.\ Pakvasa, R.J.N.\ Phillips, and K.\ Whisnant,
		Phys.\ Rev.\ D {\bf 22}, 2718 (1980).

\bibitem{Ko99}	M.\ Koike and J.\ Sato,
		Mod.\ Phys.\ Lett.\ A {\bf 14}, 1297 (1999).

\bibitem{Ot01}	T.\ Ota and J.\ Sato,
		Phys.\ Rev.\ D {\bf 63}, 093004 (2001).

\bibitem{Ja01}	B.\ Jacobsson, T.\ Ohlsson, H.\ Snellman, and W.\ Winter,
		hep-ph/0112138.

\bibitem{PREM}	A.M.\ Dziewonski and D.L.\ Anderson,
		Phys.\ Earth Planet.\ Inter.\ {\bf 25}, 297 (1981).

\bibitem{Ge01}	R.J.\ Geller and T.\ Hara, in the Proceedings of 
		{\em Nufact~'01\/} \protect\cite{NuFa}, hep-ph/0111342.

\bibitem{Li97}	E.\ Lisi and D.\ Montanino,
		Phys.\ Rev.\ D {\bf 56}, 1792 (1997).

\bibitem{Ma01}	G.L.\ Fogli, E.\ Lisi, and A.\ Marrone,
		Phys.\ Rev.\ D {\bf 64}, 093005 (2001).

\bibitem{Fo01}	G.L.\ Fogli, E.\ Lisi, and A.\ Marrone,
		Phys.\ Rev.\ D {\bf 63}, 053008 (2001).

\bibitem{Ce00}	A.\ Cervera, A.\ Donini, M.B.\ Gavela, 
		J.J.\ Gomez-Cadenas, P.\ Hernandez, O.\ Mena, 
		and S.\ Rigolin, Nucl.\ Phys.\ B {\bf 579}, 17 (2000);
		(E) {\bf 593}, 731 (2001).

\bibitem{Fr00}	M.\ Freund, M.\ Lindner, S.\ Petcov, and A.\ Romanino,
		Nucl.\ Instrum.\ Meth.\ {\bf A} 451, 18 (2000);
		M.\ Freund, P.\ Huber, and M.\ Lindner, 
		Nucl.\ Phys.\ B {\bf 615}, 331 (2001).

\bibitem{Mi01}	T.\ Miura, E.\ Takasugi, Y.\ Kuno, and M.\ Yoshimura,
		Phys.\ Rev.\ D {\bf 64}, 013002 (2001).

\bibitem{Mi02}	T.\ Miura, T.\ Shindou, E.\ Takasugi, and M.\ Yoshimura,
		Phys.\ Rev.\ D {\bf 64}, 073017 (2001).


\bibitem{KuPa}	T.K.\ Kuo and J.\ Pantaleone,
		Rev.\ Mod.\ Phys.\ {\bf 61}, 937 (1989).

		
\bibitem{Fo97}	G.L.\ Fogli, E.\ Lisi, D.\ Montanino, and G.\ Scioscia,
		Phys.\ Rev.\ D {\bf 55}, 4385 (1997).

\bibitem{Sh01}	L.\ Shan, B.\ Young, and X.\ Zhang,
		hep-ph/0110414.


\end{thebibliography}
\end{document}